\documentclass[aps,prl,twocolumn,showpacs,superscriptaddress]{revtex4}

\usepackage{graphicx,subfigure}
\usepackage{amssymb}
\usepackage{amsmath} 
\usepackage[british]{babel}
\usepackage{xcolor}

\begin{document}

\title{Distance matters: the impact of gene proximity in bacterial gene
regulation}

\author{Otto Pulkkinen}	
\affiliation{Department of Physics, Tampere University of Technology, 
FI-33101 Tampere, Finland}
\author{Ralf Metzler}
\affiliation{Department of Physics, Tampere University of Technology, 
FI-33101 Tampere, Finland}
\affiliation{Institute for Physics \& Astronomy, University of Potsdam,
D-14476 Potsdam-Golm, Germany}

\begin{abstract}
Following recent discoveries of colocalization of downstream-regulating genes
in living cells, the impact of the spatial distance between such genes on the
kinetics of gene product formation is increasingly recognized. We here show
from analytical and numerical analysis that the distance between a transcription
factor (TF) gene and its target gene drastically affects the speed and
reliability of transcriptional regulation in bacterial cells. For an explicit
model system we develop a general theory for the interactions between a TF
and a transcription unit. The observed variations in regulation efficiency are
linked to the magnitude of the variation of the TF
concentration peaks as a function of the binding site distance from the
signal source. Our results support the role of rapid binding site search
for gene colocalization and emphasize the role of local concentration
differences.
\end{abstract}


\maketitle

Suppose you live in a small town and start spreading a rumor. The time after
which the rumor reaches a specific person depends on your mutual distance,
either the physical distance due to word-of-mouth in the pre-telecommunications
era or the topological distance in modern social networks \cite{kitsak}. This 
distance dependence is immediately intuitive for random propagation in large
systems. Conversely, diffusion of signaling molecules on the 
microscopic scales of biological cells was observed to be fast
\cite{ElfLiXie}, so one might assume that spatial aspects can be neglected. Yet
recent studies strongly suggest that even in relatively small bacterial cells
distances matter with respect to both speed and reliability of genetic
regulation by DNA-binding proteins, so-called transcription factors (TFs)
\cite{Kuhlman,wunderlich}. Thus, the distance-dependence of the search time of
a given TF for its target binding site on a downstream gene was proposed to
affect the ordering of genes on the DNA, in particular, promote gene
\emph{colocalization}, i.e., the tendency of genes interacting via TFs to be
close together along the chromosome \cite{kolesov}.

\begin{figure}
\includegraphics[width=7.8cm]{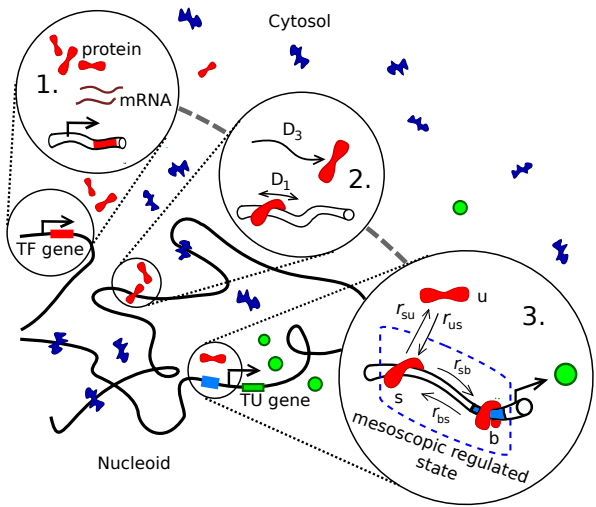}
\caption{\label{fig1} Three stochastic phases in transcriptional regulation:
\textbf{1.}~Transcription factor (TF) production. \textbf{2.}~TFs perform
facilitated diffusion in the nucleoid (inside the dashed line) containing the
DNA. Diffusion is purely 3D in the cytosol outside the nucleoid. \textbf{3.}~TFs
find  the operator of the transcription unit (TU) gene by sliding along the DNA.
The irregularly shaped blue objects depict other molecules which affect the
facilitated diffusion and binding affinity of the TF.}
\end{figure}

Transcriptional regulation, the change in gene transcription rate caused by 
binding of regulatory proteins such as TFs, is the most prominent form of 
gene regulation in bacteria \cite{Alberts}. Since TFs are proteins themselves,
their production 
consists of the inherently stochastic processes of transcription 
(conversion of the TF gene's code to RNA) and translation 
(conversion of the RNA code to proteins). Although a certain averaging of 
noise occurs due to long protein lifetimes, the noise
in the TF production propagates to downstream genes regulated by this TF
\cite{EldarElowitz:2010}. The contributions of individual stochastic steps to
the total noise in protein production (magnifying glass \textbf{1} in
Fig.~\ref{fig1}) were characterized \cite{Paulsson:2005}, and accurate
theoretical models for TF-regulated expression exist in the case of known TF
density at the regulatory site \cite{KeplerElston:2001,Friedman:2006}.

Recently, considerably effort has been invested on explaining the efficiency of
transcriptional regulation, especially the remarkable measured speed at which
TFs find their binding sites \cite{ElfLiXie,elf,gijs,riggs}. This speed is due to
facilitated diffusion \cite{bvh,mirny,coppey,bauer,klenin,marenduzzo,koslover},
in which free TF diffusion in three dimensions is interspersed by periods of
one-dimensional sliding along the DNA (Fig.~\ref{fig1}, magnifying glass
\textbf{2}). Facilitated diffusion of {\it lac} repressor molecules has indeed
been observed in living {\it E.~coli} cells \cite{elf}.
In this context, the colocalization hypothesis certainly makes sense: a shorter
search time effects more efficient regulation \cite{wunderlich}.
Concurrently, the importance of increased local protein concentration due to 
binding to DNA, occurrence of multiple binding sites, formation of protein 
complexes, and cellular compartmentalization for prokaryotic and eukaryotic 
gene regulation has been emphasized \cite{Droge:2001}.

Here we show that high local TF concentrations due to gene proximity
alone is sufficient for efficient gene regulation. Specifically, we extend
the viewpoint of TF search time optimization due to colocalization to effects
on the entire cascade of TF and TU gene expression, including the
noise in TF production, facilitated diffusion of TF, and TF binding
at TU by first binding non-specifically to the DNA and then sliding
to its specific binding site (magnifying glass \textbf{3} in Fig.~\ref{fig1}).
To our knowledge, this is the first complete, quantitative approach including
all relevant subprocesses in TF-mediated gene interaction.

The time-dependent intracellular concentration of a protein may be modeled by
stationary shot-noise \cite{Friedman:2006,Berg:1978}
$\rho(t)=V_{\mathrm{C}}^{-1}\int_{-\infty}^te^{-\gamma(t-s)} dN_B(s)$,
where $N_B$ is a compound Poisson process of protein production with combined
transcription and translation rate $a$, exponentially distributed translational
burst sizes
$B_i$, $i=0,\pm1,\pm 2,\ldots$ of mean $b$ \cite{REM_BURST}, and a combined
degradation and dilution rate $\gamma$. $V_{\mathrm{C}}$ is the (average) cell
volume. The intermediate translation step is excluded because of short mRNA
lifetimes. Under the typical fast mixing assumption of molecules in the
cell, the number of proteins
$M(\Omega,t)$ in a subdomain $\Omega$ of relative volume
$v_{\Omega}= V_{\Omega}/ V_{\mathrm{C}}$, at given time $t$,
is therefore a Poisson random variable of intensity $\int_\Omega \rho(t)
d^3{\bf r}$, with Laplace transform
\begin{eqnarray}
\label{Lapl_shot_noise}
\left< e^{-\lambda M}\right>&=&\exp\left[-a\int_{-\infty}^t\frac{b
(1-e^{-\lambda})v_\Omega e^{-\gamma(t-s)}}{1+b(1-e^{-\lambda})v_\Omega e^{-
\gamma(t-s)}}ds\right]\nonumber\\
&=&\left[1+bv_\Omega(1-e^{-\lambda})\right]^{-a/\gamma}. 
\end{eqnarray}
This is but the negative binomial distribution with parameters $a/\gamma$
and $bv_\Omega/(1+bv_\Omega)$. In particular, the mean and the variance
of the number $M(\Omega,t)$ of proteins are
\begin{equation}
\label{meanvar_ideal}
\langle M\rangle=abv_\Omega/\gamma,\,\,\,\,\,
\langle M^2\rangle-\langle M\rangle^2=abv_\Omega(1+b v_\Omega)/\gamma.
\end{equation}
Bursty protein production (large $b$) clearly effects a greater variance than
a simple Poissonian production of individual molecules. We note that the
negative binomial distribution (\ref{Lapl_shot_noise}) has been previously
found for the number of proteins in a two-stage model of stationary expression
in the fast translation limit \cite{Shahrezaei:2008}.

To study the expression of a gene controlled by a constitutive TF, we expand
the mathematical model in two ways: (i) we introduce a position dependent
kernel $\phi({\bf r},t)$ in the shot noise $\rho(t)$, to include time
delays in transcription, translation, protein folding, and, notably, facilitated
diffusion of TFs to their target site. The coordinate ${\bf r}$ is the point of
observation, namely, a point in the neighborhood of the target site (blue
operator near gene b in Fig.~\ref{fig1}, \textbf{3}). (ii) We
allow a time-dependent transcription rate $\alpha(t)$, such that the mean number
of protein production events in a time interval $[t_0,t_1]$ equals $\int_{t_0}^{
t_1}\alpha(s)ds$. The corresponding, time-inhomogeneous compound Poisson process
will be denoted by $N_{\alpha,B}$. In particular, the rate $\alpha(t)$ may be
chosen to be a \emph{random\/} process, to model fluctuations of the promoter
\cite{Paulsson:2005,Kaufmann:2007} or operator state \cite{Zon:2006},
leading to transcriptional bursts \cite{Golding-So}, see
below. The resulting process reads
$\rho({\bf r},t)=\int_{-\infty}^t\phi({\bf r},t-s)\, dN_{\alpha,B}(s)$.
Moreover, following Berg \cite{Berg:1978} instead of a continuous exponential
distribution, we will also include a discrete, geometric distribution for the
burst sizes $B$.

Even if the time evolution of the protein density $\rho({\bf r},t)$ is no
longer Markovian, we can write down its Laplace transform because, for a
given protocol $\alpha$, protein production is still a time-inhomogeneous
Poisson process:
\begin{equation}
\label{Lapl_gen_shot}
\left< e^{-\lambda\rho}|\alpha\right>=\exp\left[-\int_{-\infty}^t \!\!
\alpha(s)\frac{b\left(e^{\lambda\phi({\bf r},t-s)}-1\right)}{b\left(e^{\lambda
\phi({\bf r},t-s)}-1\right)+1}\, ds\right].
\end{equation}
The corresponding formula for the protein number is obtained by substituting
$\lambda\to1-e^{-\lambda}$ and $\phi({\bf r},t)\to\phi(\Omega,t)=\int_\Omega
\phi({\bf r},t)d^3{\bf r}$. In particular, the average of the protein number
$M(\Omega,t)$ and its variance can be immediately calculated from the Laplace
transform, yielding
\begin{subequations}
\begin{eqnarray}
\label{mean_M}
&&\hspace*{-1.2cm}
\langle M|\alpha\rangle=b\int_{-\infty}^t\alpha(s)\phi\,  ds,\\
&&\hspace*{-1.2cm}
\langle M^2|\alpha\rangle-\langle M|\alpha\rangle^2=b\int_{-\infty}^t\alpha(s)
\lbrack1+\left(2b-1\right)\phi\rbrack\phi\, ds,
\label{Var_M}
\end{eqnarray}
\end{subequations}
with $\phi=\phi(\Omega,t-s)$. Eqs.~(\ref{meanvar_ideal}) follow as
a special case of (\ref{mean_M}) and (\ref{Var_M}) with a 
constant transcription rate, large burst size, 
and infinitely fast mixing of molecules in a homogeneous cell volume,
i.e., $\phi(\Omega,t)=v_{\Omega}e^{-\gamma t}$.

Let us now consider the effect of a TF (here, a repressor) to the transcription
rate $\alpha_{\mathrm{TU}}$ of a transcription unit (TU) gene under control of
the TF. We first assume a given density of unbound TF within the sliding
distance along the DNA from the operator site, and study the local kinetics
of the TF. We explicitly describe the local kinetics of the repressor molecules
through facilitated diffusion \cite{bvh,Zon:2006} near the binding site by
considering three states of the operator (magnifying glass \textbf{3} in
Fig.~\ref{fig1}): transcription occurs at a constant rate $a$ whenever there
is no repressor bound to the DNA at the target. Then, the repressor
is either performing a local search by sliding in the vicinity of the target
without binding to it specifically, or TF molecules, the mean number of which
is determined by the given density, are just hovering in the surrounding space.
The gene is considered silent when a repressor is bound at the operator. The
linear Markov dynamics of TF binding can be explicitly solved by standard
methods (see Supplementary Material (SM) \cite{sm}). For example, the stationary
protein level is obtained by averaging over $\alpha$ in Eq.~(\ref{mean_M}), but
its variance will consist of three terms instead of the two in Eq.~(\ref{Var_M})
because of time correlations in the transcription rate $\alpha$.

Introducing the equilibrium constant $K_{\mathrm{SP}}$ for specific TF binding
to the operator and assuming fast binding and unbinding, we integrate out the
fast local search state in the three-state model.This leads to a simpler model 
with telegraph noise at the operator, i.e., the gene is either silent or being
transcribed at some effective rate $a_{\mathrm{eff}}$. The transitions between
these two states occur without intermediates at rate $r_{\mathrm{on}}$ from
silent to active and \emph{vice versa\/} with rate $r_{\mathrm{off}}$. Matching
the stationary mean and the variance of the protein numbers in both processes,
we relate the parameters of the telegraph model to the ones depicted in the
magnifying glass \textbf{3} of Fig.~\ref{fig1} \cite{sm}. This is the description of a
mesoscopic repressed state discussed in Ref.~\cite{Zon:2006}, where it is argued
that this choice of retaining the completely silent state in the coarse-grained
theory is justified by the separation of timescales in local search dynamics
and RNA polymerase (RNAP) binding; the rebinding of repressor is extremely
fast, thus leaving hardly any time for RNAP to intervene \cite{Zon:2006}. Of
course, there exists another telegraph scenario that would leave the original
transcription rate for the completely unbound state untouched, but would
introduce an effective, \emph{leaky\/} transcription rate for the combined
repressed state consisting of nonspecifically and specifically bound states.
This alternative scenario is certainly plausible. For example, the leaky
expression of \emph{lac\/} genes \cite{Yu:2006}, has been associated with
DNA looping \cite{Choi:2008}. We
do not consider this point further here.

We now address the interaction of TF and TU genes via repression and study
the transient response of the TU gene to a change in the transcription rate of
the TF gene when the latter is turned on at $t=0$ and then constitutively
expressed. We study the dynamics of the moments of the TU gene transcription
rate $\alpha_{\mathrm{TU}}$ as functions of the distance between the genes.
From simulated
trajectories (Fig.~S1 \cite{sm}) of suitably normalized repressor
concentrations (see below) within a binding distance from the target and the
resulting expression levels of the gene under control, the TF shows distinct
concentration peaks for a pair of vicinal genes, and a fast decrement in
expression level of the TU gene due to TF binding.

To analytically model the TF searching its binding site, we assume a linear
dependence of the nonspecific binding rate on the repressor concentration near
the target and introduce the equilibrium non-specific binding constant $K_{
\mathrm{NS}}$. If the basal rate, in absence of repressors, of expression of
the TU gene is $a_{\mathrm{TU}}$, the mean and variance of the transcription
rate $\alpha_{\mathrm{TU}}(r,t)$ under repression become
\begin{subequations}
\begin{eqnarray}
\label{Ealpha} 
&&\hspace*{-0.6cm}
\langle\alpha_{\mathrm{TU}}\rangle=\langle a_{\mathrm{eff}}\frac{r_{
\mathrm{on}}}{r_{\mathrm{on}}+r_{\mathrm{off}}}\rangle=a_{\mathrm{TU}}p_{
\mathrm{on}}(r,t),\\
\label{Varalpha}
&&\hspace*{-0.6cm}
\langle\alpha_{\mathrm{TU}}^2\rangle-\langle\alpha_{\mathrm{TU}}\rangle^2=
a_{\mathrm{TU}}^2p_{\mathrm{on}}(r,t)\left[1-p_{\mathrm{on}}(r,t)\right],
\end{eqnarray}
where we use the probability that the TU gene is actively transcribed at time
$t$ when the gene-gene distance is $r$,
\begin{equation}
\label{ponE}
p_{\mathrm{on}}(r,t)=\left<\frac{1+K_{\mathrm{NS}}\rho_{\mathrm{TF}}(\Omega,t)}
{1+K_{\mathrm{NS}}(1+K_{\mathrm{SP}})\rho_{\mathrm{TF}}(\Omega,t)}\right>.
\end{equation}
\end{subequations}
As a typical example, $\Omega$ is a tube surrounding the sliding region around
the target. Its length is 34 nm (100 base pairs), its diameter is
that of DNA (2.4 nm) plus 30 nm ({\it e.g.}, the length of {\it lac}
repressor is 14 nm). With Eq.~(\ref{Lapl_gen_shot}),
\begin{equation}
\label{pon}
p_{\mathrm{on}}(r,t)=\frac{1}{1+K_{\mathrm{SP}}}\left(1+\int_0^{\infty}e^{
-\lambda-\int_{-\infty}^t \alpha_{\mathrm{TF}}(s) \frac{\aleph}{1+\aleph}ds}
d\lambda\right),
\end{equation}
where $\aleph=b_{\mathrm{
TF}}(\exp\{\lambda\tilde{K}\phi(\Omega,t-s)\}-1)$ and $\tilde{K}=(1+K_{\mathrm{
SP}})K_{\mathrm{NS}}$. The lower limit of the inner integral can be set to zero
in our scenario ($\alpha_{\mathrm{TF}}(t<0)=0$).

Eq.~(\ref{pon}) is a central result of this study. It is general and allows
quantitative analyses of various transcriptional and translational repression
scenarios in any cellular structure and geometry. In
particular, it takes into account the transciptional pulsing \cite{Golding-So}
of the TU gene induced by the binding of the repressor. Eq.~(\ref{pon}) even
allows us to model RNAP binding and mRNA degradation by setting $b_{\mathrm{TF}}
=1$ and introducing, as the TF production rate, a new stochastic process
$\alpha_{\mathrm{
TF}}(t)=v_{\mathrm{TF}}N_{\mathrm{mRNA}}(t)$ with a constant translation rate
$v_{\mathrm{TF}}$, and the number of transcripts $N_{\mathrm{mRNA}}$ given
by an immigration-death process (equivalently an $\mathrm{M}/\mathrm{M}/\infty$
queue) with mRNA production rate $a_{\mathrm{TF}}$ and mRNA degradation rate
$\gamma_{\mathrm{mRNA}}$. Since $\gamma_{\mathrm{mRNA}}$ is of the same order
as typical TF search times in \emph{E.~coli} \cite{ElfLiXie,elf}, inclusion of
TF mRNA dynamics may be necessary in some cases. The scenario
can be even further extended to include TF transcriptional pulsing by modulating
the immigration-death process $N_{\mathrm{mRNA}}$ with telegraph noise
\cite{Kaufmann:2007}. However, the expectation of Eq.~(\ref{pon}) is yet to be
solved for these $\alpha_{\mathrm{TF}}$ \cite{Crawford}.
In the examples below, we use an approximation with a constant
transcription and translation rate yielding on average 500 TF molecules per
cell under stationary conditions. This number is in the
ballpark of TF abundances for various levels of {\it E. coli\/}
regulation networks \cite{Janga:2009}. Special cases with low and high TF
abundances will be studied separately.

We assume the TF gene to be in the
center of a spherical nucleoid and the TU gene at a radial distance $r$ from
it. There is recent evidence \cite{Kuhlman} that the spatial distribution of
TFs is highly inhomogenous. TFs bind to the DNA nonspecifically, hence under
many growth conditions the TF concentration is higher in the nucleoid than
in the surrounding volume. Inhomogeneities were also observed to affect fold
repression. We thus assume that the diffusion constant $D_\mathrm{N}$
within the nucleoid is much smaller than in the surrounding cytosol due to
crowding and nonspecific binding to the DNA (see SM \cite{sm} for comparison
with the model in Ref.~\cite{Kuhlman}). 
The nucleoid is surrounded by the volume $V_{\mathrm{C}}-
V_{\mathrm{N}}$, where $V_{\mathrm{C}}=4\pi R_{\mathrm{C}}^3/3= 1$ $\mu\mathrm{
m}^3$ and $V_{\mathrm{N}}=4\pi R_{\mathrm{N}}^3 /3=0.2$ $\mu\mathrm{m}^3$ are
the cell and nucleoid volumes. The density $\rho_{\mathrm{TF}}(r,t)$ is subject
to the radial diffusion equation. In Eq.~(\ref{pon}), $\phi(\Omega,t)\approx
V_{\Omega}\phi(r,t)$ obeys
\begin{eqnarray}
\nonumber
&&\frac{\partial\phi}{\partial t}=D_{\mathrm{N}}\left(\frac{\partial^2\phi}{
\partial r^2}+\frac{2}{r}\frac{\partial\phi}{\partial r}\right)-\gamma\phi,
\mbox{ for }0\leq r\leq R_{\mathrm{N}}\\
\label{diffeq2}
&&\frac{\partial\phi}{\partial t}=-\frac{4\pi R_{\mathrm{N}}^2
D_{\mathrm{N}}}{V_{\mathrm{C}}-V_{\mathrm{N}}}\frac{\partial\phi}{\partial r}
-\gamma\phi,\mbox{ for }r=R_{\mathrm{N}},
\end{eqnarray}
with a dilution rate $\gamma = 1/20$ $\mathrm{min}^{-1}$ due to cell growth and
with the condition that the TFs are initially uniformly distributed in
the close vicinity (say, within a radius $R_{\mathrm{I}}=20$ nm) of the TF
gene. This is justified from the observed localization of transcripts near their
transcription site in bacteria \cite{Llopis:2010}. The explicit solution of
Eqs.~({\ref{diffeq2}}) for our spherical geometry is \cite{CarslawJaeger:Book}
\begin{eqnarray}
\nonumber
\phi(r,t)&=&\frac{e^{-\gamma t}}{V_{\mathrm{C}}}+\frac{3}{2\pi}
\sum_{n = 1}^{\infty}e^{-(D_{\mathrm{N}}q_n^2+\gamma)t}\frac{\sin
(q_nr)}{R_{\mathrm{N}}r}\times\\
&&\hspace*{-2.0cm}
\times\frac{k^2\psi_n^4+3(2k+3)\psi_n^2+9}{k^2 \psi_n^4+9(k+1)\psi_n^2}\cdot \frac{\sin(\theta_n)
-\theta_n\cos(\theta_n)}{R_{\mathrm{I}}\theta_n^2},
\label{phi}
\end{eqnarray}
where $\psi_n= q_n R_{\mathrm{N}} $, $\theta_n=q_nR_{\mathrm{I}}$ and $k=(V_{\mathrm{C}}-V_{\mathrm{N}})/V_{
\mathrm{N}}$, and the $q_n$ are the positive solutions of $(3+kR_{\mathrm{I}}^2
q^2)\tan(qR_{\mathrm{I}})=3qR_{\mathrm{I}}$. Eq.~(\ref{phi})
is our other central result. 

Fig.~2 shows the probabilities (\ref{pon}) as function of time
for short and long distances between the TF and TU genes. Accordingly,
the distance impacts vastly the regulation
efficiency: the response is significantly stronger and faster for short
distances, this difference persisting for minutes.
Fig.~2 also demonstrates that it is necessary to consider this exact expression
instead of a mean field approximation obtained by taking expectations of the
density separately in the numerator and denominator in Eq.~(\ref{ponE}). The
mean field approximation would overestimate the spatial differences
in regulation. Therefore, it is of importance to use the exact formula
(\ref{pon}) instead.

\begin{figure}
\includegraphics[height=7.2cm,angle=270]{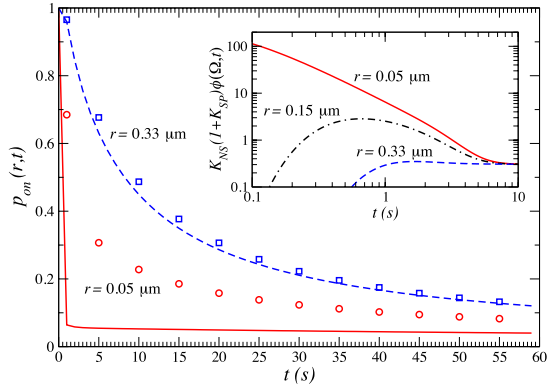}
\caption{\label{fig2} Transient response to a change in TF transcription rate.
The circles and squares are the probabilities (\ref{pon}) for TF-TU gene
distances $r=0.05$ and $0.33$ $\mu$m, and the solid and dashed lines show the
corresponding mean field approximations (see main text). The inset shows the
variation of TF concentration around the target site at various TF-TU distances.
The equilibrium constants are $K_{NS}=10$, $K_{SP}=1000$, and the rest of the
parameters as described in the main text.}
\end{figure}

The inset of Fig.~2 shows the reason for the difference between
exact and mean field approaches: as already suggested
by the simulated trajectories in Fig.~S1, the amplitude variation
of the TF concentration contributing to nonspecific binding at the target
depends heavily on the separation of TF and TU genes. The TU
genes far away from the TF gene receive a more diluted signal than those
close-by. Specifically, both Fig.~S1 and the inset
of Fig.~2 show $\tilde{K}\phi(\Omega,t)$,
characterizing both the availability of TF and its binding affinity to the
target. Its values should be compared to 1, the scale set by the first
term in the exponential of Eq.~(\ref{pon}). The truncation of the peak
observed at short distances causes the mean field theory to fail. Note that
smaller TF copy numbers than used
here lead to a similar spatial effect in $p_{on}$; e.g., the
same set of parameters but with a stationary mean number of 100 TFs
leads to a roughly constant difference of the order $0.1$ between $p_{on}$
with $r=0.33$ and $0.05$ $\mu$m in a window of 1 min. The magnitude of the
effect depends naturally on the TF binding affinity at the target. Both the
expression levels and binding specificity are known to depend on whether the TF
is a local or global regulator \cite{Lozada-Chavez:2008, Janga:2009}.

With Eq.~(\ref{Varalpha}), we assess the noise propagation
in the TF-TU system, in particular, the variance of the transcription rate of
the TU gene. Since the variances are proportional to the product $p_{on}(r,t)
\left[1-p_{on}(r,t)\right]$, we see from Fig.~2 that they peak at a few seconds
and at ten seconds for $r=0.05$ and $0.33$ $\mu$m, respectively. The
probability $p_{on}$ grows with distance to the TU gene, and the same hence
applies to the variance after the initial transient peak. The
total time-integrated variance is greater for the distant gene, and its
transcription is therefore more susceptible to stochastic variation in TF
production. However, the effect in Eq.~(\ref{Varalpha}) is small 
for small $\alpha_{\mathrm{TU}}$, and the situation may be different under 
stationary conditions. Fig.~2 shows that the distance variation in expression 
levels in the long time limit can be small, even if the transient response 
shows considerable variation. The same applies to expression fluctuations. 
Experimental observations \cite{Salman-Taniguchi} show that the protein level
fluctuations are, in general, determined by the mean expression level, 
and are independent of system details. The dependence of protein number
fluctuations on the TF-TU distance under stationary conditions needs to be
explored further.

Concluding, we established a quantitative model for the distance dependence
of gene regulation efficiency and stochasticity in bacteria. Intracellular
structure and nonspecific binding to the DNA are taken into account in terms of
an inhomogeneous diffusion rate. The binding at the target is facilitated by a
local search process, which was modeled by an intermediate fast degree of
freedom. Significant spatial effects in the regulation efficiency was
demonstrated, strongly supporting the regulation hypothesis for gene
colocalization. We note that more precise models, for instance, with multiple
TFs sliding simultaneously near the target can be solved, as well. The
expressions are more elaborate (except for infinite numbers) but the binding
probabilities show roughly the same behavior as above. It will be of interest
to compare transient response to internal and external signals, as the gene
location is known to depend on the type of signal \cite{Janga:2007}.

We acknowledge funding from the Academy of Finland (FiDiPro scheme).


\begin{thebibliography}{21}
\bibitem{REM_BURST} A translational burst refers to the number of times the transcripts (mRNA)
are translated into proteins.
\end{thebibliography}
\end{document}